\documentclass[12pt]{iopart}
\usepackage{graphicx}
\begin{document}

\letter{Spanning trees for the geometry and dynamics of compact polymers}

\author{Armin Rahmani, Andrea Velenich and Claudio Chamon}

\address{Department of Physics, Boston University, Boston,
MA 02215 USA}

\date{\today}

\begin{abstract}
Using a mapping of compact polymers on the Manhattan lattice to spanning trees, we calculate exactly the average number of bends at infinite temperature. We then find, in a high temperature approximation, the energy of the system as a function of bending rigidity and polymer elasticity. We identify the universal mechanism for the relaxation of compact polymers and then endow the model with physically motivated dynamics in the convenient framework of the trees. We find aging and domain coarsening after quenches in temperature. We explain the slow dynamics in terms of the geometrical interconnections between the energy and the dynamics. 

\end{abstract}


\maketitle

\textsl{
Introduction:}
Polymers confined to spaces much smaller than their radius of gyration are common in biological systems such as DNA packed in viral capsids~\cite{cerritelli97} or in the nucleus of a cell~\cite{Kleckner04}. Both the geometrical structure and the dynamical behavior of strongly confined polymers are of considerable interest because of their role in biological functions. One of the first theoretical models for such phases is Flory's \textit{compact polymer model} (CPM)~\cite{Flory56} which  describes a polymer by a Hamiltonian walk on a lattice. This two-dimensional model is relevant for polymers on surfaces~\cite{Katzav06} such as DNA adsorbed on a lipid bilayer. The thermodynamics of Flory compact polymers on the square lattice has been studied using field theoretical methods~\cite{Jacobsen98,Jacobsen04}. Little is known, however, about the geometrical structure of compact polymers. Moreover, despite the fact that the Flory's CPM was introduced more than fifty years ago, the dynamics of the model has not been studied so far.

In this letter, we use a mapping of the compact polymer on the Manhattan lattice to spanning trees~\cite{Kasteleyn63, Duplantier88} to address the problems above.
This mapping allows us to obtain an exact analytical expression for the average number of bends and then calculate the energy of the system in a high temperature approximation. More importantly, we uncover a deep universal connection between the dynamics of compact polymers and their geometric structure. We identify the extension and retraction of geometric features called \textit{fingers} as the main mechanism for the relaxation of compact polymers. We then use the mapping to the spanning trees to endow the Flory's model with physically motivated kinetic rules. We find aging and slow relaxation in the quench dynamics of the system and argue that the slow relaxation stems from the geometrical correlations between the energy and the number of fingers. The relaxation phenomenology presented in this paper gives a universal description of polymer dynamics under strong geometric confinement.

\begin{figure}
\centerline{
\includegraphics[width=6.0 cm] {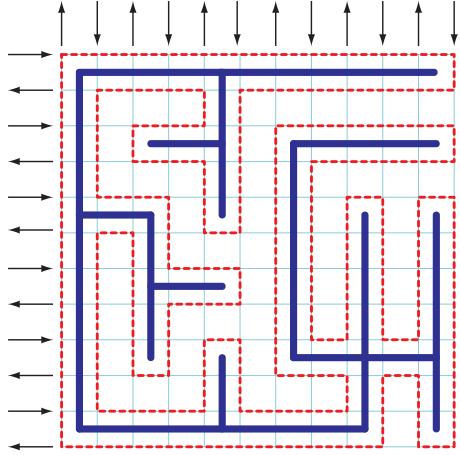}
}
\caption{(Color online) A compact loop polymer (dashed red line) and the corresponding spanning tree (bold blue line).
}
\label{fig:mapping}
\end{figure}

\begin{figure}
\centerline{
\includegraphics[width=12.0 cm] {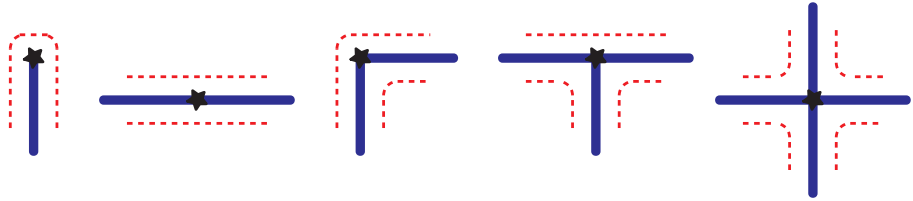}
}
\caption{(Color online) Configuration of the tree (bold blue line) and the polymer (dashed red line) at sites (black stars) with coordinations $1, 2a, 2b, 3$ and $4$.
}
\label{fig:site}
\end{figure}

\begin{figure}
\centerline{
\includegraphics[width=4.0 cm] {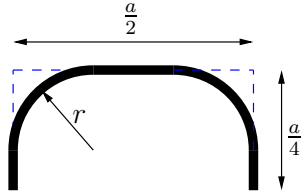}
}
\caption{Length fluctuation at a site with coordination $1$.
}
\label{fig:cap}
\end{figure}

\textsl{Model and Thermodynamics:}
The \textit{compact} polymer is modeled as a Hamiltonian cycle, i.e.~a self-avoiding loop which visits every site, on the Manhattan lattice (a square lattice with successive parallel lines oriented in alternating directions) with lattice spacing $a/2$. As seen in Fig.~\ref{fig:mapping}, there is a one-to-one correspondence between polymer configurations and spanning trees on a square lattice with lattice spacing $a$. A spanning tree is a connected graph that visits every site of this new lattice without forming loops.
Hereafter we work with the lattice underlying the trees and assume it has $N=m \times m$ sites. In a tree, a vertex can have coordination $1$, $2$, $3$ and $4$; sites with coordination $2$ can have the two bonds either ``aligned'' (coordination $2a$) or ``bent'' (coordination $2b$). Denoting the number of sites with coordination $\alpha$ by $N_\alpha$, we have
\begin{eqnarray}\label{constraint}
\nonumber
&&  N_1+N_2+N_3+N_4=N \\
&&  N_1+2N_2+3N_3+4N_4=2N
\end{eqnarray}
We assume periodic boundary conditions so there are $2N$ bonds on the lattice with $(m-1)(m+1)\simeq N$ bonds belonging to the tree.
For a given polymer configuration, we consider the bending energy and the elastic energy due to length fluctuations. As seen in Fig.~\ref{fig:site}, vertices of the tree with coordination $1$, $2a$, $2b$, $3$ and $4$ correspond to the polymer having $2$, $0$, $2$, $2$ and $4$ bends respectively.
Hence using Eq.~(\ref{constraint}), the total number of bends, is given by
\begin{eqnarray}\label{bends}
N_B=2(N+N_4-N_{2a})
\end{eqnarray}
If the characteristic energy of one bend is $\varepsilon_b$, the bending energy is given by $ E_b = \varepsilon_b N_B$. If we consider a polymer whose unstretched length is twice the length of the tree, and assume that the bends are rounded with a radius $r$, we can show using elementary geometry that the sites with coordination $1$, $2a$, $2b$, $3$ and $4$ contribute to the length fluctuation by $a-4r+\pi r$, $0$, $-4r+\pi r$, $- a-4r+\pi r$ and $-2a-8r+2\pi r$ respectively. For example, the length fluctuation for a site with coordination $1$ is equal to the length of the curve shown in Fig.~\ref{fig:cap}.
Using the above fluctuations and Eq.~(\ref{constraint}), it can be shown that the total length fluctuation depends only on $N_B$ and is given by $\frac{1}{2}(\pi-4)rN_B$. By adding the corresponding elastic energy to the bending energy, we get for the total energy of the system
\begin{equation}\label{Energy}
    E(N_B)=\varepsilon_b N_B+\frac{\kappa a}{16N}(\pi-4)^2(\frac{r}{a})^2{N_B}^2
\end{equation}
where the spring constant of the polymer is $\frac{\kappa}{2Na}$. To find the average number of bends $\langle N_B \rangle$, we need $p_4$ and $p_{2a}$ where $p_\alpha$ is the probability of a site belonging to the tree having coordination $\alpha$ .

\paragraph{}
The probabilities $p_1$, $p_2$, $p_3$ and $p_4$ were calculated in reference~\cite{Manna92} using Kirchhoff's matrix-tree theorem~\cite{Kirchhoff1847}. Notice that because of constraints in Eq.~(\ref{constraint}), only two of these four probabilities are independent.
 \begin{equation}\label{p}
p_1=\frac{8}{\pi^2}-\frac{16}{\pi^3}, \qquad p_4=-1+\frac{8}{\pi}-\frac{20}{\pi^2}+\frac{16}{\pi^3}
 \end{equation}
We also need $p_{2a}$, which we calculate below. Kirchhoff's theorem states that the number of spanning trees on a graph is given by any cofactor (they are all equal) of its Laplacian matrix. The Laplacian matrix of a graph with $N$ vertices is an $N$ by $N$ matrix whose diagonal elements $\Delta_{ii}$ are the coordination number of vertex $i$ and the off-diagonal elements $\Delta_{ij}$ are $-1$ times the number of links between vertices $i$ and $j$. Representing empty, occupied and unconstrained bonds by \begin{picture}(15,5) \put(0,2){\includegraphics[width=0.03\textwidth]{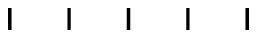}}\end{picture} , \begin{picture}(15,5) \put(0,2){\includegraphics[width=0.03\textwidth]{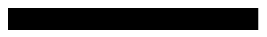}}\end{picture} and \begin{picture}(15,5)\put(0,2){\includegraphics[width=0.03\textwidth]{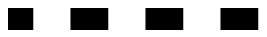}}
\end{picture} respectively, we can write
\begin{equation}\label{relation}
 p\left( \begin{picture}(22,0)\put(2,-6){\includegraphics[width=0.035\textwidth]{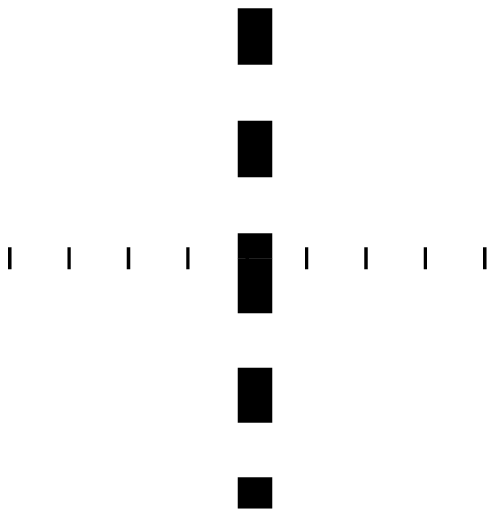}}\end{picture} \right) = p\left( \begin{picture}(22,0)\put(2,-6){\includegraphics[width=0.035\textwidth]{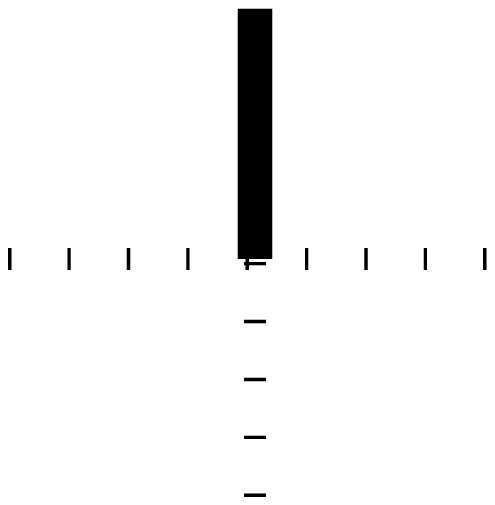}}\end{picture} \right) + p\left( \begin{picture}(22,0)\put(2,-6){\includegraphics[width=0.035\textwidth]{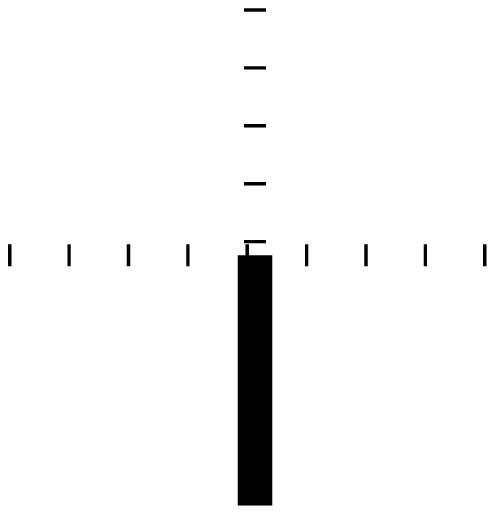}}\end{picture} \right) + p\left( \begin{picture}(22,0)\put(2,-6){\includegraphics[width=0.035\textwidth]{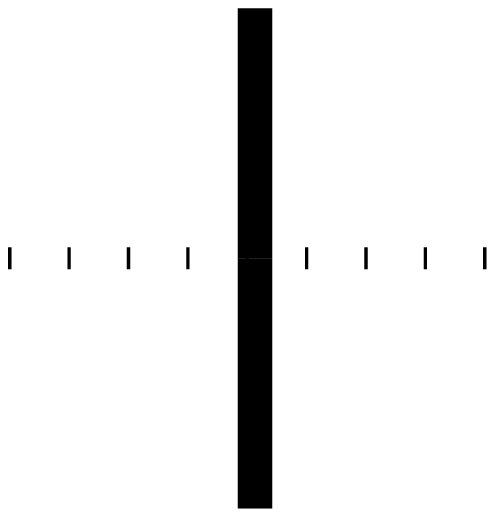}}\end{picture} \right)
\end{equation}
where $p(\cdots)$ is the probability of the four bonds emanating from a site having the shown configuration. The first and second terms in the right hand side of Eq.~\ref{relation} are each equal to $p_1/4$ and the last term to $p_{2a}/2$. The left hand side of Eq.~\ref{relation}, $p\left( \begin{picture}(22,0)\put(2,-6){\includegraphics[width=0.035\textwidth]{prob1.eps}}\end{picture} \right)\equiv p_c$ is equal to ratio of the number of spanning trees on a square lattice with two adjacent horizontal bonds removed and the total number of spanning trees on the original square lattice. Denoting the associated Laplacian matrices by $\Delta+B$ (the lattice minus two bonds) and $\Delta$ (the lattice), we have according to the matrix tree theorem $p_c=\frac{\textrm{cof}(\Delta+B)}{\textrm{cof}(\Delta)}$. Using the identity $\det (XY ) = \det X \det Y$, the ratio of cofactors can be written as ${\rm det}(I + GB)$ where $G(r-r^\prime)=\Delta^{-1}(r,r^\prime)-\Delta^{-1}(r^\prime,r^\prime)$ is the regularized inverse Laplacian. Noticing that the Laplacian of the square lattice is diagonal in Fourier space we have ~\cite{Majumdar91}
 \begin{equation}\label{G}
    \begin{array}{ll}
      G(\pm a\hat{x})=G(\pm a\hat{y})=-\frac{1}{4} & G(\pm a\hat{x}\pm a\hat{y})=-\frac{1}{\pi} \\
      G(\pm 2a\hat{x})=G(\pm 2a\hat{y})=-1+\frac{2}{\pi} &
    \end{array}
 \end{equation}
Since the matrix $B$ describes a change in the connectivity of $3$ sites, it only has a $3 \times 3$ nonzero block. Then $p_c=\det(I+GB)$ can be calculated by evaluating the determinant of a $3 \times 3$ matrix. We can then obtain $p_{2a}$ using $p_1$ from Eq.~\ref{p} and the result is $p_{2a}=2p_c-p_1=\frac{4}{\pi}-\frac{16}{\pi^2}+\frac{16}{\pi^3}$, in the limit of an infinite lattice.
Thus in the thermodynamic limit, the average number of bends in a closed compact polymer $\langle N_B \rangle=2N(1+p_4-p_{2a})$ can be finally calculated and is given by
\begin{equation}\label{Nbend}
 \langle N_B \rangle=N(\frac{8}{\pi}-\frac{8}{\pi^2}).
\end{equation}

\paragraph{}
The above probabilities have been computed assuming all configurations are equally likely (infinite temperature). For finite high temperatures the energy of the system can be obtained as follows: for $\alpha=1,2a,2b,3,4$, let us define $f_{\alpha}(x)$ as the probability distribution function that a tree configuration has $N_{\alpha}=x$. Then $f_{\alpha}(x)$ has a peak at $Np_\alpha$ and we have numerically validated a Gaussian approximation for the fluctuations around the mean value of $Np_\alpha$ with standard deviations scaling as $\sigma \sim \sqrt{N}$ by generating random spanning trees using Broder’s algorithm~\cite{Broder89}. Similarly the distribution function of $N_B$ is approximated by a Gaussian peaked at $\langle N_B \rangle$ given by
\begin{equation}\label{f_M}
    f(N_B)=\frac{1}{\sqrt{2\pi N\sigma_0^{2}}} e^{-\frac{(N_B-\langle N_B \rangle)^2}{2N\sigma_0^{2}}}
\end{equation}
where $\sigma_0^{2}=0.645\pm0.003$ has been obtained numerically.
Within the Gaussian approximation~(\ref{f_M}), the partition function $Z\propto\int_{-\infty}^{\infty}e^{\gamma N}f(N_B)e^{-\beta E(N_B)} dN_B$ where the number of spanning trees on a square lattice with $N$ sites asymptotically scales as $e^{\gamma N}$~\cite{Wu77} with $\gamma=\frac{4}{\pi}(1-\frac{1}{3^2}+\frac{1}{5^2}-\frac{1}{7^2}+...)$. From the partition function, the average energy of the system in the thermodynamic limit and at inverse temperature $\beta$ is obtained as
\begin{equation}\label{energy}
    E(\beta)=N\frac{(m_0-\beta \varepsilon_b \sigma_0^{2})( m_0 \varepsilon_e +\varepsilon_b + \beta \varepsilon_e \varepsilon_b \sigma_0^{2})}{(1+2\beta \varepsilon_e \sigma_0^{2})^2}
\end{equation}
where $m_0=\frac{\langle N_B \rangle}{N}$ and $\varepsilon_e=\frac{\kappa a}{16}(\pi-4)^2(\frac{r}{a})^2$ is a characteristic elastic energy. In the absence of length fluctuations, the energy reduces to $E(\beta)=N\varepsilon_b(m_0-\beta \varepsilon_b \sigma_0^{2})$. 
 Notice however that because of the high-temperature nature of the approximation, this equation does not hold for arbitrarily large $\beta \varepsilon_b$ (the total energy cannot be negative) and it does not capture the melting transition the system~\cite{Jacobsen04}.

\begin{figure}
\centerline{
\includegraphics[width=8.0 cm] {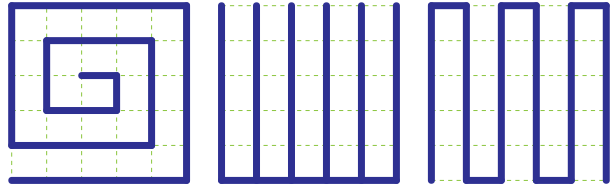}
}
\caption{
Some zero temperature configurations of the spanning tree which minimize the bending energy.
}
\label{fig:ground}
\end{figure}

\paragraph{}
The system has a melting transition~\cite{Jacobsen04} since the polymer is disordered at infinite temperature and the $T=0$ configurations are ordered as seen in Fig.~\ref{fig:ground}.
To find the transition temperature, we define the phenomenological order parameter $D_{\ell}=(N_v-N_h)/(N_v+N_h)$, with $N_v$ and $N_h$ the number of vertical and horizontal bonds in a macroscopic (but not too large) region of size $\ell$. Calculating the Binder cumulant $B_{\ell}=1-\frac{\langle D_{\ell}^4\rangle}{3\langle D_{\ell}^2\rangle^2}$~\cite{Binder81} for $\ell=20,28, 36, 44$ to identify the transition we obtain $\beta_c \simeq 0.8$. We calculated the cumulant using an ensemble average with $256$ realizations obtained by Monte-Carlo (MC) simulations. Furthermore, the specific heat obtained numerically has a peak close to the transition whose finite size scaling indicates no thermodynamic limit divergence. This is suggestive of a Kosterlitz-Thouless transition ~\cite{Berker79}. The transition is analogous to that of ~\cite{Jacobsen04} where the compact polymer lives on the square rather than the Manhattan lattice.

\begin{figure}[h]
\centerline{
\includegraphics[width=10.0 cm] {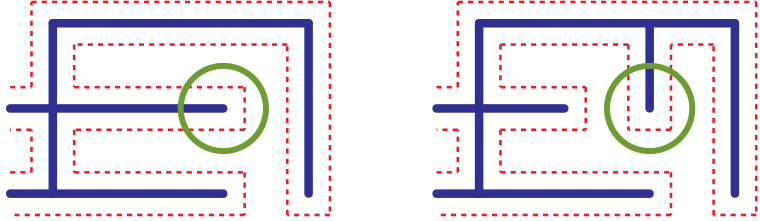}
}
\caption{
A fingering event.
}
\label{fig:finger}
\end{figure}

\textsl{Quench Dynamics and Aging:}
In this section we study the dynamics of a compact polymer. Notice that for a single strongly-confined polymer, the center of mass does not diffuse and the system dynamics is characterized only by reshaping. In order to study the dynamics of the system, we need to endow the Flory's model with some kinetic rules. These should represent the actual motion of the polymer rather than just transform the configurations as needed for thermodynamic MC studies. In reference~\cite{Rahmani07}, we studied the dynamics of a confined single polymer using a microscopic model. We found that the reshaping of strongly confined polymers takes place primarily through \textit{fingering events} in which a finger retracts and another finger extends to fill the void (Fig.~\ref{fig:finger}). 
In the language of the trees, a fingering event is a local bond flip at a site with coordination one: a bond is added to one of the three empty edges while the original bond emanating from the coordination-one site is removed from the tree. Since a fingering event involves the collective motion of many monomers, this dynamics is greatly accelerated.
For a given move the change in the system energy $\Delta E$ depends only on the change in the coordination of the sites involved in the bond flip. To satisfy detailed balance we use Glauber-like dynamics where the rate $r$ of a flip is proportional to $e^{-\beta \Delta E}$. As explained in reference~\cite{Rahmani07}, although the shape of the polymer is modified locally, the actual polymer undergoes global motion through reptation~\cite{deGennes71}. Therefore, in addition to the energetics, a kinetic factor controls the rate at which the moves take place.
To find the kinetic factor in the language of the trees, consider one fingering event. In a closed polymer there are two paths for monomers to flow from the retracting finger to the extending one; both paths contribute to the motion but in the thermodynamic limit, reptation essentially takes place along the shorter path of length $\ell_s$. In order to create a new finger, the excess length coming from the retracting finger diffuses along the shorter path of length $\ell_s$ with a diffusion constant proportional to ${1 \over \ell_s}$~\cite{DoiEdwards}. Since the polymer reptates a fixed distance for each fingering event, we have $r \propto{1 \over \ell_s}$. Adding a bond to the tree (representing the extending finger) creates one and only one loop. This loop partitions the system into two regions of areas $s=S$ and $s=N-S$, each containing one of the two polymer paths. For a path lying inside a region of area $s$, the length of the reptating polymer is then $l=p-2+2(s-p/2+1)=2s$ where $p$ is the perimeter of the region and $(s-p/2+1)$ is the number of lattice sites inside it.
The rate $r$ will then be proportional to $\frac{1}{\textrm{min}(S,N-S)}$.
The dynamics we have introduced above is ``non-local'' in the sense that for every move a non-local quantity $S$ needs to be computed. A faster, ``local'' dynamics can be implemented by substituting $\frac{1}{s}$ with the infinite-temperature average $\langle\frac{1}{s}\rangle$. The same thermodynamics and similar dynamical features as for the non-local dynamics are obtained.

\begin{figure}[h]
\centerline{
\includegraphics[width=12.0 cm] {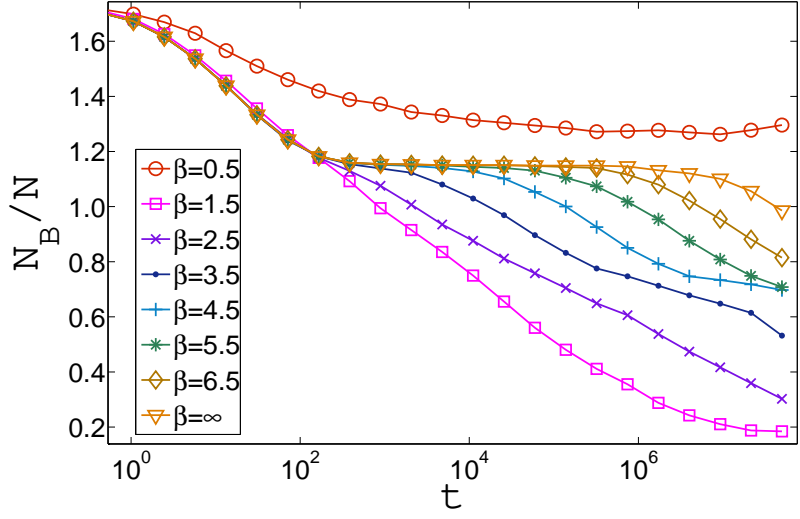}
}
\caption{
Emergence of a non-equilibrium plateau in the quench dynamics of the system.
}
\label{fig:dynamics}
\end{figure}

With open boundary conditions, we can show that, in the thermodynamic limit, the region with smaller area is the \textit{inside} of the loop formed by adding a bond. At infinite temperature, this is a consequence of that fact the probability of forming loops of area $s$ decays asymptotically as $s^{-\frac{11}{8}}$~\cite{Manna92}. At low temperatures, the typical loop area $S$ scales as the linear size of the ordered domains which is much smaller than $N-S$.
\paragraph{}
Using both the local and the non-local dynamics, we study the response of an ensemble of spanning trees to a sudden quench from infinite temperature to a non-vanishing inverse temperature $\beta$. In the simulations we only consider the bending rigidity and neglect the elastic energy; temperatures are measured in natural units where $\varepsilon_b=1$. We measure how the average $N_B\propto E$ and $N_1$ decay as a function of time. Since the dynamics takes place only through fingering events at the leaves of the tree (the mobile nodes), $N_1$ is related to the system mobility. As seen in Fig.~\ref{fig:dynamics}, for small values of $\beta$ the energy decays steadily to its equilibrium value whereas for larger $\beta$ a broad plateau emerges. $N_1$ has a similar behavior. All plateaus emerge at approximately the same value of $\frac{N_B}{N}$ and $\frac{N_1}{N}$ for different temperatures. The results presented are for $N=28^2$. A parallel study for $N=18^2$ indicates no system size dependence up to the time-scales we studied. For quenches to temperatures which allow the system to equilibrate within our simulation times, the system size independence of the relaxation time holds during the whole evolution. For quenches to lower temperatures, we observe that the time evolution is the same for the two different system sizes until the early stages of the secondary relaxation. The time scale necessary to reach complete ordering at such low temperatures is indeed expected to depend on the system size but such time scales are not accessible to our simulations. 

\begin{figure}[h]
\centerline{
\includegraphics[width=18.0 cm] {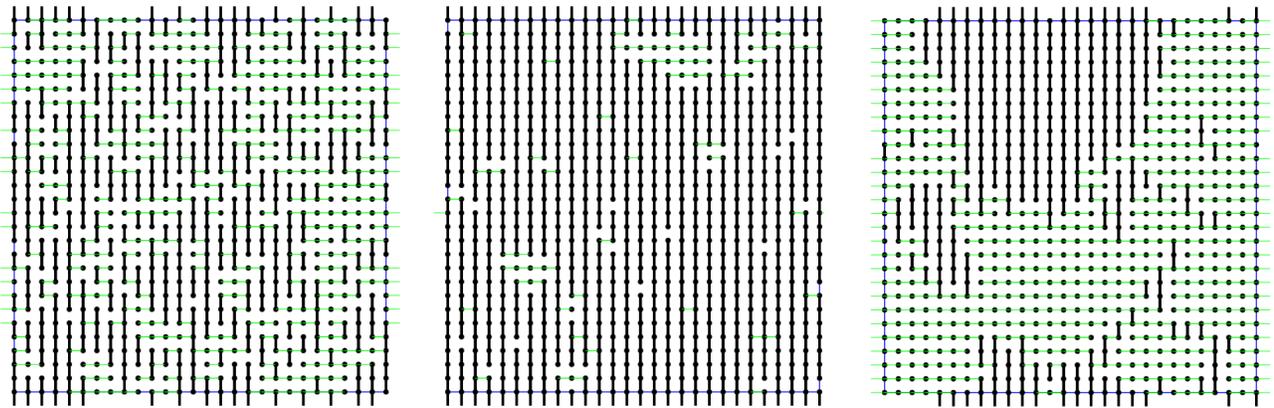}
}
\caption{
(Color online: thick black vertical bonds, thin green horizontal bonds) Left: a disordered equilibrium state after a quench from $\beta_i=0$ to $\beta_f=0.5<\beta_c$. Center: an ordered equilibrium state for $\beta_f=1.0>\beta_c$. Right: coexistence of domains in a non-equilibrium plateau for a quench to $\beta_f=3.0$.
}
\label{fig:visualization}
\end{figure}

The dynamical features we observe can be explained by noting that none of the moves which decrease the energy increase the number of mobile nodes $N_1$. Hence, as the system decreases its energy, the average distance between mobile nodes grows, freezing larger and larger parts of the system. Also, mobile nodes can be created or annihilated only if they are nearest neighbor to another mobile node and they can diffuse only by hopping over another mobile node. At zero temperature, only the moves that \textit{do not increase} the energy are allowed and the plateau emerges once the moves that \textit{decrease} the energy are exhausted. While in the plateau, the system is exploring its phase space with the moves that \textit{do not change} the energy and a secondary decay begins only once the cooperative reshaping due to such moves has made further energy-decreasing moves accessible. At finite low temperatures the behavior is similar, although the finite probability of energy-increasing moves shortens the plateau. A relaxation mechanism governed by fingers and the slow dynamics due to the geometric correlation between the number of fingers and the system energy are generic dynamical features which do not depend on the details of the model. Notably, with the non-local dynamics, $N_B$ and $N_1$ decay into a plateau of the same height as for the local dynamics. The decay, however, happens more slowly since the average loop area controlling the decay rate increases with time.  
\paragraph{}
\begin{figure}[h]
\centerline{
\includegraphics[width=12.0 cm] {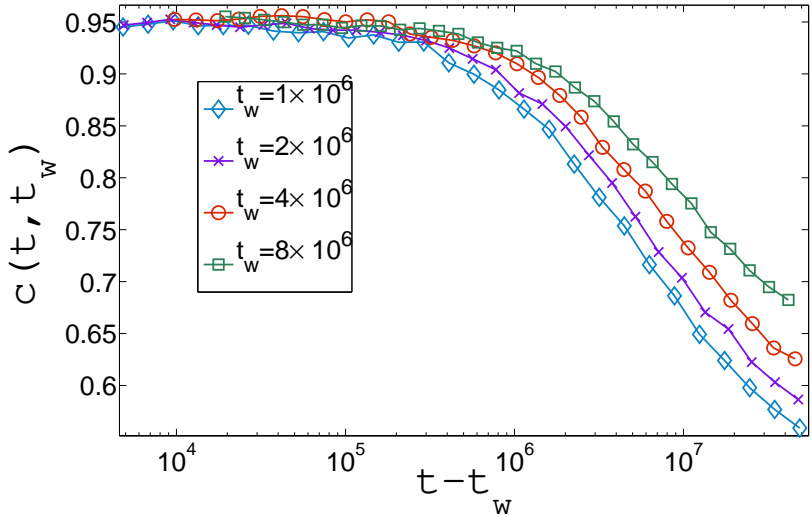}
}
\caption{
Aging in the shape-shape correlation function $c(t,t_w)$ for $\beta=6.5$.
}
\label{fig:aging}
\end{figure}
We now define a shape-shape correlation function for the system $c(t,t_w)=\frac{1}{N}\sum_{i}^{N}\delta_i(t,t_w)$,
where $\delta_i(t,t_w)=1$ if the configuration of bonds emanating from site $i$ are exactly the same at times $t_w$ and $t$ and zero otherwise. As seen in Fig.~\ref{fig:aging}, the system decorrelates more slowly for longer waiting times $t_w$. The aging in the system is characterized by domain coarsening. As seen in Fig.~\ref{fig:visualization}, the ordered equilibrium state below the critical temperature consists primarily of a domain of horizontal or vertical bonds. When the system is driven out of equilibrium several ordered domains persist for long times (plateau) and undergo a slow coarsening during the secondary decay (aging).

\paragraph{}
\textsl{Summary and Outlook:}
We used a mapping to spanning trees to calculate quantities related to the geometrical structure of Flory's compact polymers.  We found that the dynamics of compact polymers is closely related to their geometric structure due to the role of fingers in reshaping. Formulating the dynamics in terms of the collective polymer reshaping, we endowed the model with physically motivated kinetics in the convenient framework of the trees. Our accelerated dynamics gives access to the long-time out-of-equilibrium phenomenology of compact polymers. We found that the system exhibits slow dynamics and aging characterized by the coarsening of ordered domains when quenched to low temperatures. We presented a generic phenomenological picture for the dynamics of strongly confined polymers by noting that the key to the slow relaxation is the geometrical correlations between the number of fingers and the system energy. Even in the continuum, where fingers do not have a rigorous definition, fingering events are a generic mechanism driving the relaxation dynamics of polymers~\cite{Rahmani07} and qualitative dynamical features such as domain coarsening and aging found in our paper are not expected to depend on the details of the lattice model. Quantitative results like the number of bends however are specific to the lattice studied in the paper.  Our model could be extended to study a fully-packed melt of loop polymers instead of single chain by using a mapping to a spanning forest and appropriate kinetic rules. Similar studies in the context of non-compact polymers have been done on a related problem of chains diffusing in an array of obstacles~\cite{Obukhov94,Cates86,Nachaev87}.

The simulations were carried out on Boston University supercomputer facilities. We are grateful to G. Biroli, C. Castelnovo, J. Jacobsen, J. Kondev, A. Meller, S. Redner,  J. Schmit, and A. Sportiello for helpful discussions and correspondence. This work was supported in part by the NSF Grant DMR-0403997.


\section*{References}
\begin{thebibliography}{99}

\bibitem{cerritelli97} M.E. Cerritelli {\em et al.},
    \emph{Cell}  {\bf 91}, 271-280 (1997).

\bibitem{Kleckner04}
N. Kleckner {\em et al.}, Proc. Natl. Acad. Sci. USA \textbf{101},
12592-12597 (2004).

\bibitem{Flory56}
P. J. Flory, Proc R. Soc. London A \textbf{234}, 60 (1956).

\bibitem{Katzav06}
E. Katzav, M. Adda-Bedia and A. Boudaoud, Proc. Natl. Acad. Sci. USA \textbf{103},
18900-18904 (2006).

\bibitem{Jacobsen98}
J. L. Jacobsen, J, Kondev, Nucl. Phys. B \textbf{532}, 635 (1998).

\bibitem{Jacobsen04}
J. L. Jacobsen, J. Kondev, Phys. Rev. Lett. \textbf{92}, 210601 (2004).

\bibitem{Kasteleyn63}
P. W. Kasteleyn, Physica \textbf{29}, 1329 (1963).

\bibitem{Duplantier88}
B. Duplantier, F. David, J. Stat. Phys. \textbf{51}, 327 (1988).


\bibitem{Manna92}
S. S. Manna, D. Dhar and S. N. Majumdar, Phys. Rev. A \textbf{46}, R4471 (1992).

\bibitem{Kirchhoff1847}
G. Kirchhoff, Ann. Phys. Chem. \textbf{72}, 497 (1847).

\bibitem{Majumdar91}
S. N. Majumdar and D. Dhar, J. Phys. A: Math. Gen. \textbf{24}, L357-L362 (1991).

\bibitem{Broder89}
A. Z. Broder, in \emph{Proceedings of the 30th annual IEEE Symposium on Foundations of Computer Science,}
(IEEE, New York, 1989), p.442

\bibitem{Wu77}
F. Y. Wu, J. Phys. A: Math. Gen. \textbf{10}, L113-L115 (1977).

\bibitem{Binder81}
K. Binder, Z. Phys. B: Condens. Matter \textbf{43}, 119 (1981).

\bibitem{Berker79}
A. N. Berker and D. R. Nelson, Phys. Rev. B \textbf{19}, 2488 (1979).

\bibitem{Rahmani07}
A. Rahmani, C. Castelnovo, J. Schmit, C. Chamon,
J. Stat. Mech. P09022 (2007).

\bibitem{deGennes71}
P. G. de Gennes, J. Chem. Phys. \textbf{55}, 572 (1971) .

\bibitem{DoiEdwards}
M. Doi and S. F. Edwards, The Theory of Polymer Dynamics, Oxford University Press, (1986).

\bibitem{Obukhov94}
S. P. Obukhov, M. Rubinstein, T. Duke, Phys. Rev. Lett. \textbf{73}, 1263 (1994).

\bibitem{Cates86}
M. E. Cates and J. M. Deutsch, J. Physique  \textbf{47} 2121 (1986).

\bibitem{Nachaev87}
S. K. Nechaev,  A. N. Semenov and M. K. Koleva, Physica A \textbf{140} 506 (1987).

\end{thebibliography}
\end{document}